\documentclass[aps,prb,twocolumn,showpacs,amsmath,amssymb,superscriptaddress]{revtex4-1}

\usepackage{graphicx}
\usepackage{dcolumn}
\usepackage{bm}

\usepackage{color}
\usepackage{ulem}
\begin{document}
\preprint{APS/123-QED}


\title{
Microscopic mechanisms of initial formation process of graphene on SiC(0001) surfaces}

\author{Fumihiro Imoto}
\affiliation{Department of Applied Physics, The University of Tokyo, Hongo, Tokyo 113-8656, Japan}
\author{Jun-Ichi Iwata}
\affiliation{Department of Applied Physics, The University of Tokyo, Hongo, Tokyo 113-8656, Japan}
\author{Mauro Boero}
\affiliation{Institut de Physique et Chimie des Mat{\' e}riaux de Strasbourg UMR 7504,
University of Strasbourg and CNRS, 23 rue du Loess, F-67034 Strasbourg, France}
\author{Atsushi Oshiyama}
\affiliation{Department of Applied Physics, The University of Tokyo, Hongo, Tokyo 113-8656, Japan}

\date{\today}

\begin{abstract}
We report total-energy calculations based on the density-functional theory that clarify microscopic mechanisms of initial stage of graphene formation on the SiC(0001) surface. We explore favorable reactions for desorption of either Si or C atoms from the stepped surface by determining the desorption and the subsequent migration pathways and calculating the corresponding energy barriers for the first time. We find that the energy barrier for the desorption of an Si atom at the step edge and the subsequent migration toward stable terrace sites are lower than that of a C atom by 0.75 eV, indicative of the selective desorption of Si from the SiC surface. We also find that the subsequent Si desorption is an exothermic reaction. This exothermicity comes from the energy gain due to the bond formation of C atoms being left near the step edges. This is certainly a seed of graphene flakes. 

\end{abstract}
\maketitle


\section{introduction}\label{intro}

Graphene is a target of an intense forefront research both in fundamental science and in nano-technological applications because of its peculiar electronic and structural properties. Specifically, its relatively large surface area and rich edge termination possibilities make graphene particularly appealing for nanoelectronics\cite{geim}, biosensors\cite{shang}, energy storage and conversion\cite{shao}, and nanocatalysis\cite{yu,ikeda}. In the specific field of semiconductor devices, the high carrier mobility and the possibility of patterning or growing nanographene flakes of desired shape and size\cite{mayorov,bolotin,masubuchi,tedesco,wu,li} have paved the way to new possibilities in the efforts of overcoming limitations of both current and future electronics.

The originally proposed exfoliation of graphite \cite{geim} is still a practical and useful way to make high-quality graphene. However, 
its poor yield makes this method unsuitable for technological applications. An alternative and promising way to form graphene is the thermal decomposition of silicon carbide (SiC) \cite{mishra}, since it can find direct applications in the realization of electronic devices. SiC is a semiconductor widely used in power electronics because of its robustness in harsh environments such as high voltage and high temperature \cite{ieee}. Thermal decomposition, upon proper thermodynamical conditions, is a versatile technique to dislodge Si atoms from the SiC network, leaving behind high quality graphene flakes or islands\cite{mishra}. This thermal decomposition is promising in a wealth of applications since it provides large-area graphene supported on regular SiC substrates. Being SiC a wide band-gap semiconductor, the resulting graphene/SiC heterostructure is particularly appealing for direct applications in electronic devices. Yet, the process that leads to the formation of this type of supported nanographene has still to be fully unraveled. In fact, such a reaction involves drastic modifications of the bonding environment and related electronic structure that are beyond the reach of experimental probes.

What is known to date is that Si atoms are desorbed selectively from the SiC, leading to a carbon-rich $6\sqrt{3}\times6\sqrt{3}R30^{\circ}$ buffer layer \cite{hannon08,kim_ihm}, and then eventually graphene layers are formed on the SiC(0001) surface. Electron microscopy experiments have revealed that the formation of these graphene layers occurs where atomic steps are present on the SiC surface \cite{norimatsu10} and, particularly, in the regions where the step density is high \cite{hannon08}. In fact, the SiC (0001) exhibits peculiar nanofacets \cite{kimoto,nakamura,fujii,arima,nie,nakagawa} resulting upon heat treatment and subsequent hydrogen etching done to polish the surface \cite{norimatsu10,riedl}. This type of nanofacet is now identified as a bunched single bilayer atomic steps, for which atomistic calculations \cite{sawada1,sawada2} based on the density-functional theory (DFT) \cite{kohn1,kohn2} have been done aimed at inspecting the local structure. After the formation of the buffer carbon layer, subsequent graphene layers can grow from the interface of the buffer layer and SiC\cite{hannon11}. 
However, the underlying mechanism is still elusive.

To unravel the microscopic mechanisms responsible for this formation of graphene structures on SiC, we resort to the DFT--based calculations aimed at clarifying the pathways along which Si atoms are dislodged from the SiC network. We work out the associated energy barriers characterizing the possible reaction paths, tracking the processes that lead to the formation of the initial carbon seeds from which graphene can subsequently grow. Former DFT calculations for carbon layers on SiC(0001) \cite{kageshima09,kageshima11,kageshima13} were focused mainly on the structural stability of a selected set of carbon overlayers and, as such, they provide little information about the actual reaction mechanisms. From the experimental observation described above, the essential and basic processes of the graphene growth seem to be summarizable in two stages: (i) Si atoms are desorbed from the nanofacet, i.e., the bunched atomic steps, and (ii) the aggregation of residual C occurs near the step edges from which the Si atoms departed. Starting our investigation from this basic idea, we elucidate the mechanisms of the elemental processes responsible for the formation of the initial graphene seeds and provide a comprehensive picture of all possible reaction pathways that the system can exhibit upon desorption, including the fate of the Si atoms that leave the SiC network. The desorption processes on which we focus here are the ones in which an edge atom is dislodged and, subsequently, it migrates to suitable sites on 
either a lower or an upper terrace. The choice of these specific atoms is clearly dictated by the aforementioned experimental evidence\cite{norimatsu10,hannon08}.

In this paper, we consider all possible Si and C desorption pathways from nanofacets in order to get a comprehensive scenario about the desorption of both chemical species and the possible edge sites from which such desorption might occur. We then consider the migration of either Si or C desorbed atom to the sites called T4 or H3 which have been identified as stable adatom sites on the terrace\cite{northrup}.

The organization of the present paper is as follows. In section \ref{method}, we explain the methodology and computational approaches used in the present set of calculations. The obtained reaction pathways for the desorption of Si and C from the step edge and their subsequent migration are presented and discussed in section \ref{results}. Finally, we summarize our findings in section \ref{sum}.

\section{Computational Methods}\label{method}

SiC exists as various polytypes in which the stacking of atomic bilayers along the bond direction differs. 
Each one of these polytypes is identified and labeled according to the periodicity of the stacking sequence and its symmetry (cubic or hexagonal) such as 2H (wurtzite), 3C (zincblende), 4H, 6H and so forth. Among all these possible polytypes, the one labeled as 4H is the most stable, being characterized by the larger cohesive energy and, for this reason, commonly used in experiments. The sequence of the biatomic layers along the (0001) direction in 4H-SiC is $\cdot \cdot$ ABCB $\cdot \cdot$. The cleavage of this bulk structure can result in two inequivalent surfaces: In one surface, starting from the exposed top layer, the sequence is ABCB$\cdot \cdot$ and in the other it is BCBA $\cdot \cdot$. The former is referred to as cubic surface, whereas the latter is termed hexagonal surface, following the name of the bulk polytype with the same atomic-layer sequence. 
\begin{figure}
\includegraphics[width=1.0\linewidth]{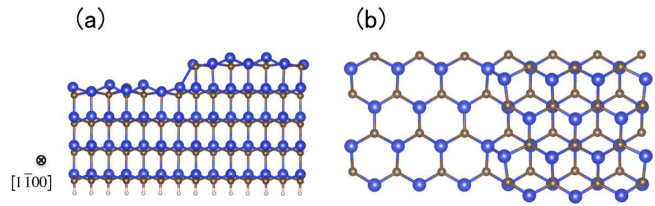}
\caption{(color online)
(a) Side and (b) top views of the optimized geometry of the SB step model used in the present work. 
Blue, brown, and white spheres indicate Si, C, and H atoms, respectively.}\label{init-c-edge}
\end{figure}
As mentioned in section \ref{intro}, graphene is formed, at least during the initial stage, from the surface steps of the 
(11$\bar{2}$n) nanofacets ($ n \approx$ 12) 
on SiC(0001) surfaces \cite{norimatsu10}. Microscopically, the nanofacet is composed of bunched single bilayer (SB) steps \cite{sawada1,sawada2}. In this work, we consider the desorption of either Si or C atoms from SB steps of SiC(0001) surfaces and the subsequent migration of the desorbed atom. The system is modeled as a periodic supercell containing one slab in which the top surface presents upper and lower terraces bordered by a SB step parallel to the $[1\bar{1}00]$ direction. Our model has a lateral size of $7\times2\sqrt{3}$ and is composed of 280 atoms. The bottom surface is fixed to the bulk crystallographic positions and terminated by H atoms to compensate for the missing bonds, whereas the rest of the system is allowed to evolve and relax freely. Each slab consists of five or four bilayers (see, e.g., Fig.~\ref{init-c-edge}) and each slab is separated by a vacuum region of more than 8 {\AA} to minimize the interaction between its periodically repeated images. 

All simulations are done with the RSDFT code\cite{RSDFT1,RSDFT2,RSDFT3}, in which the Kohn-Sham equation is discretized on a three-dimensional grid in real space. The real-space grid spacing is set to be 0.21 {\AA}, corresponding in the reciprocal Fourier space to an energy cut--off of 62.3 Ry. For the exchange--correlation functional, the generalized gradient approximation of Perdew, Burke and Ernzerhof\cite{PBE} is adopted. The core--valence interaction is described by norm-conserving pseudopotentials generated by the recipe proposed by Troullier and Martins\cite{TM1,TM2}. The Brillouin-zone integration is performed on two $k$-points sampled along the step edge direction. All geometry optimizations are done under the convergence criterion of residual forces $< 1.0\times10^{-3}$ Hartree/a.u (i.~e. 51.4 meV/{\AA}). The geometry optimized SB step structure constructed in this way is shown in Figure \ref{init-c-edge}. 
The equilibrium distance between the Si atom on the step edge and the adjacent Si site on the lower terrace is 2.43 {\AA}, considerably shorter than the unrelaxed length (3.10 {\AA}) as cleaved from the bulk, but longer than the typical Si--Si bulk bond length (2.37 {\AA}). 

Reaction pathways and energy barriers are calculated by using the hyperplane constraint method\cite{jeong_oshiyama}. In this method, we first determine a line in the 3$N_{\rm atom}$ dimensional space ($N_{\rm atom}$: the number of atoms in the simulation cell) which connects the initial and the final or the intermediate metastable structures for a selected reaction pathway. Then a suitable number of points along these lines is selected and geometry optimizations are performed on each point by minimizing all the force components perpendicular to this line: i.e., the constraint minimization on each hyperplane perpendicular to the line. This results in probable reaction pathways. It is necessary to select sufficiently large number of the points to ensure a good and fine sampling of the pathway.

In this paper, we present the results for the SB step where the top-most surface is cubic for the upper terrace and hexagonal for the lower one, as shown, e.g., in Fig.~\ref{init-c-edge}.  
An alternative geometry for the SB step consists in a hexagonal upper terrace and a cubic lower terrace.
We have also examined the desorption from this alternative system and found that the energetics in the desorption processes are analogous to that from the SB step with the cubic upper and the hexagonal lower terraces, as will be explained in the following  section.

\begin{figure}
\includegraphics[width=.98\linewidth]{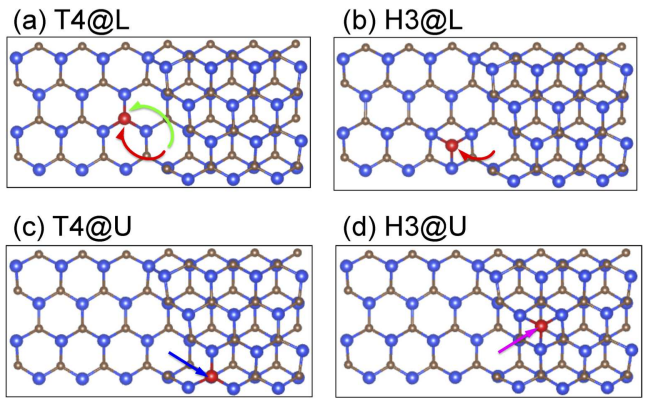}
\caption{(color on line)
Top view of the optimized final structures in which an edge Si atom 
is desorbed and migrates 
to T4 or H3 sites. In the four panels, 
the upper and lower terraces are shown in the right and left parts, respectively. One 
Si atom is desorbed to a lower T4 site (a), a lower H3 site (b), an upper T4 
site (c), and an upper H3 site (d).
Arrows in each panel schematically indicate the reaction paths followed by the dislodged atom. 
The color of each arrow is consistent with the color code used in Fig.~\ref{1stSi-dE}.
}\label{fin-Si}
\end{figure}
\begin{figure}
\includegraphics[width=.98\linewidth]{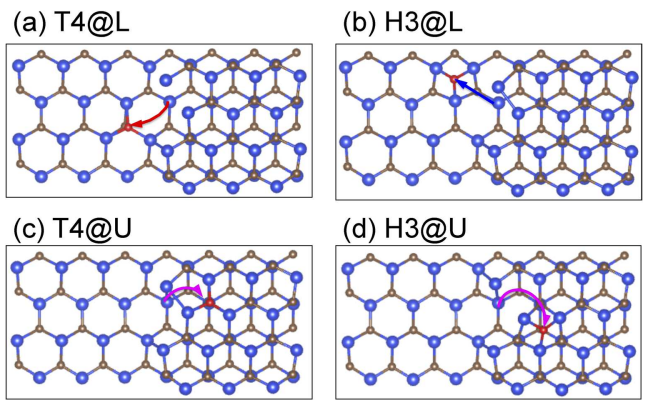}
\caption{(color on line)
Top view of the optimized final configurations in which an edge C atom 
is desorbed to T4 or H3 sites.
In the four panels, 
the upper and lower terraces are shown in the right and left parts, respectively. One 
C atom is desorbed to a lower T4 site (a), a lower H3 site (b), an upper T4 site (c),
and an upper H3 site (d).
Arrows in each panel schematically indicate the reaction paths followed by the dislodged atom. 
The color of each arrow is consistent with the color code used in Fig.~\ref{1stC-dE}.
}\label{fin-C}
\end{figure}

\section{results and discussion}\label{results}

\subsection{First desorption from the step edge}\label{ssec:1Si}

To cope with the complexity of the possible processes along which the desorption of atoms from a step edge 
can evolve, we have considered all the possible alternative pathways for either Si or C departing from their 
original position and ending to a stable T4 or H3 site located on the terrace. The final configurations 
for these reaction pathways are shown in Figs.~\ref{fin-Si} and \ref{fin-C}. 
In each figure, the desorbed Si or C atom is indicated by a red sphere.
Across the whole discussion, we use the notation T4(or H3)@L and T4(or H3)@U to indicate the reaction
in which an atom is desorbed and migrates to a T4(or H3) site on the lower (@L) and upper (@U) terraces, respectively. 
\begin{figure}
\includegraphics[width=1.0\linewidth]{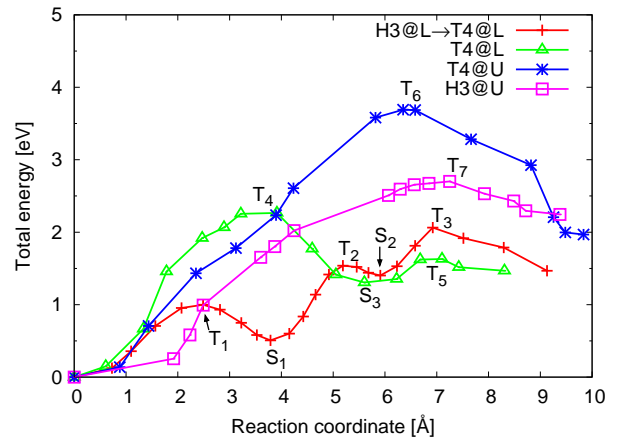}
\caption{
Calculated energy barriers of Si desorption from the step edge to T4 and H3 sites on the terrace, as indicated by the upper right legend. $S_i (i = 1, \cdot \cdot 3)$ and $T_i ( i=1, \cdot \cdot, 7)$ are metastable and transition states, respectively, in the reactions. $S_{2}$ is of the same geometry as H3@L in Fig.~\ref{fin-Si}(b). The color of each reaction path corresponds to that of each arrow in Fig.~\ref{fin-Si}.
}
\label{1stSi-dE}
\end{figure}

A first result can be summarized as follows: Whenever an edge Si atom is desorbed and migrates onto a T4 or an H3 site, all the possible final configurations shown in the four panels of Fig.~\ref{fin-Si} are less stable than the initial structure. This lower stability is far from negligible, since the four structures (a), (b), (c), and (d) are energetically located above the initial configuration by 1.46, 1.40, 1.95 and 2.26 eV, respectively. The same picture holds also for the cases in which an edge C atom is desorbed and migrates onto a T4 or an H3 site. These structures correspond to the panels (a), (b), (c), and (d) of Fig.~\ref{fin-C} and are characterized by the energy increases of 1.44, 1.05, 0.89 and 0.45 eV from the initial configuration. Each reaction is therefore endothermic, being the final product higher in energy than the initial state. For a SB step with the hexagonal upper and cubic lower terraces, the energetics among the initial and the final structures are essentially identical: Each final geometry also has higher energy than the initial one; the energy increases are 1.13 eV (T4@L), 1.11 eV (H3@L), 2.07 eV (T4@U), 2.20 eV (H3@U) for Si, whereas 0.70 eV (T4@L), 1.68 eV (H3@L), 0.56 eV (T4@U), 0.48 eV (H3@U) for C.

\begin{figure}
\includegraphics[width=.98\linewidth]{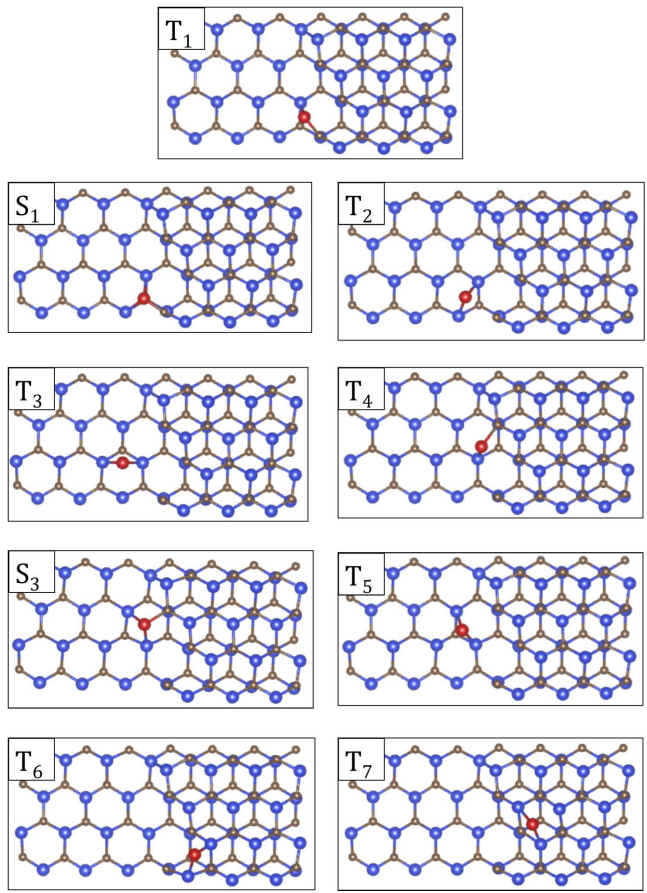}
\caption{
Metastable ($S_i$) and transition-state ($T_i$) geometries along the total energy profiles of Fig.~\ref{1stSi-dE} in which an edge Si is desorbed onto the T4 or the H3 site on the terrace. The moving Si is evidenced by a large red sphere in each panel. 
}
\label{1stSi-S-T}
\end{figure}

Figure \ref{1stSi-dE} shows the total energy profiles for the reactions in which an edge Si atom is desorbed and migrates to T4 and H3 sites on the upper and lower terraces. In determining each reaction pathway, we have defined several lines which connect the initial, final or intermediate (meta)stable configurations in the 3$N_{\rm atom}$ dimensional space, defined the hyperplanes, and performed the constraint minimization. The abscissa in Fig.~\ref{1stSi-dE} is the distance in the 3$N_{\rm atom}$ dimensional space from the initial configuration along thus determined reaction pathway. The ordinate
indicates the total energy with respect to the energy of the initial configuration, namely the same SB step structure.

We have found several metastable geometries $S_i (i=1, \cdot \cdot 3)$ and transition-state geometries $T_i (i=1, \cdot \cdot, 7)$ along the reaction pathways as shown in Fig.~\ref{1stSi-dE}. 
These geometries are shown in Fig.~\ref{1stSi-S-T}. The largest energy difference between the metastable and transition-state geometries represents the main barrier to be overcome and, as such, the rate-determining step for each selected reaction.

In the process in which the edge Si is desorbed and migrates to the site of the lower terrace, we have identified two different pathways: i.e., in one pathway (red crosses in Fig.~\ref{1stSi-dE}) the Si migrates first to the H3@L site and then the T4@L site, and in the other way (green triangles in Fig.~\ref{1stSi-dE}) it migrates directly to the T4@L site. In the first pathway, we have found two metastable geometries $S_1$ and $S_2$ and the three transition states, $T_1$, $T_2$ and finally $T_3$ toward the T4@L site.
In $T_{1}$, the Si-C bond between the departing 
Si and the edge-C is cleaved. This energetically demanding 
electronic-structure
rearrangement is characterized by the relatively large barrier of 1.00 eV. We remark that $S_{1}$ is a local minimum, in which the 
desorbed 
Si is bonded to 
C atoms at the edge and at a lower T4 site. From $S_{1}$, the system changes to a metastable structure $S_{2}$ (Fig.~\ref{1stSi-dE}), that is identical to the final state of H3@L [Fig.~\ref{fin-Si}~(b)], overcoming a transition state $T_{2}$ with the rate-determining barrier of 1.03 eV. From $S_{2}$, 
the Si overcomes a small barrier of 0.60 eV at the transition state $T_3$ and eventually reaches the T4@L site.
In the direct pathway to the
T4@L site, the Si is first desorbed to a metastable site as shown in $S_{3}$ after overcoming the rate-determining barrier of 2.27 eV at the transition state $T_{4}$. Then, from $S_{3}$, this same Si atom migrates to the T4@L site via a transition state $T_{5}$ characterized by a small barrier of 0.32 eV.

In the process in which an edge Si atom is desorbed and migrates to the sites at the upper terrace, we have identified two pathways. The Si reaches the T4@U and the H3@U after overcoming the transition states $T_6$ and $T_7$, respectively (blue asterisks and pink squares in Fig.~\ref{1stSi-dE}). 
The calculated energy barriers are 3.69 eV and 2.70 eV, respectively. We have found that the departing 
Si becomes twofold coordinated in each transition geometry, whereas it is threefold coordinated in each (meta)stable configuration.

\begin{figure}
\includegraphics[width=1.0\linewidth]{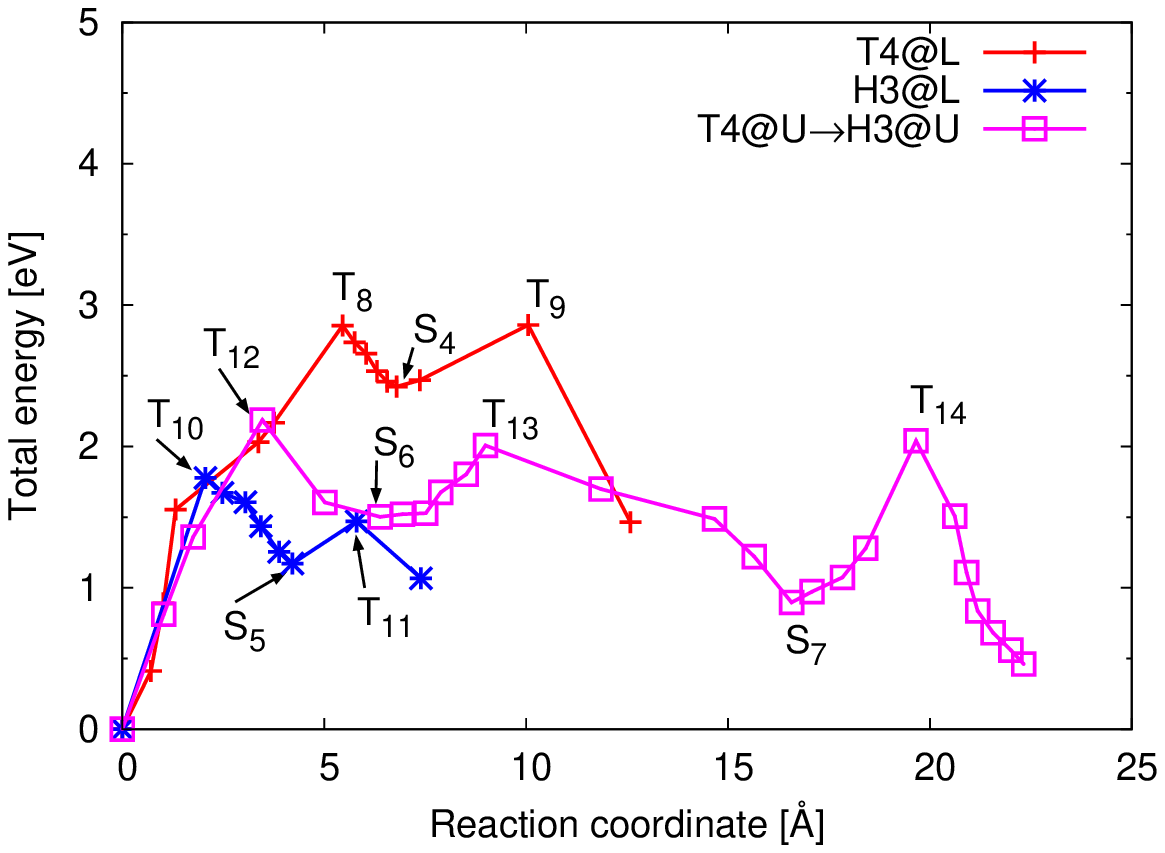}
\caption{
Calculated energy barriers of C desorption from the step edge to T4 and H3 sites on the terrace, as indicated by the upper right legend. $S_i (i = 4, \cdot \cdot 7) $ and $T_i ( i=8, \cdot \cdot, 14)$ are metastable and transition states, respectively, in the reactions. $S_{7}$ is of the same geometry as T4@U in Fig.~\ref{fin-C}(c). The color of each reaction path corresponds to that of each arrow in Fig.~\ref{fin-C}. 
}
\label{1stC-dE}
\end{figure}
\begin{figure}
\includegraphics[width=.98\linewidth]{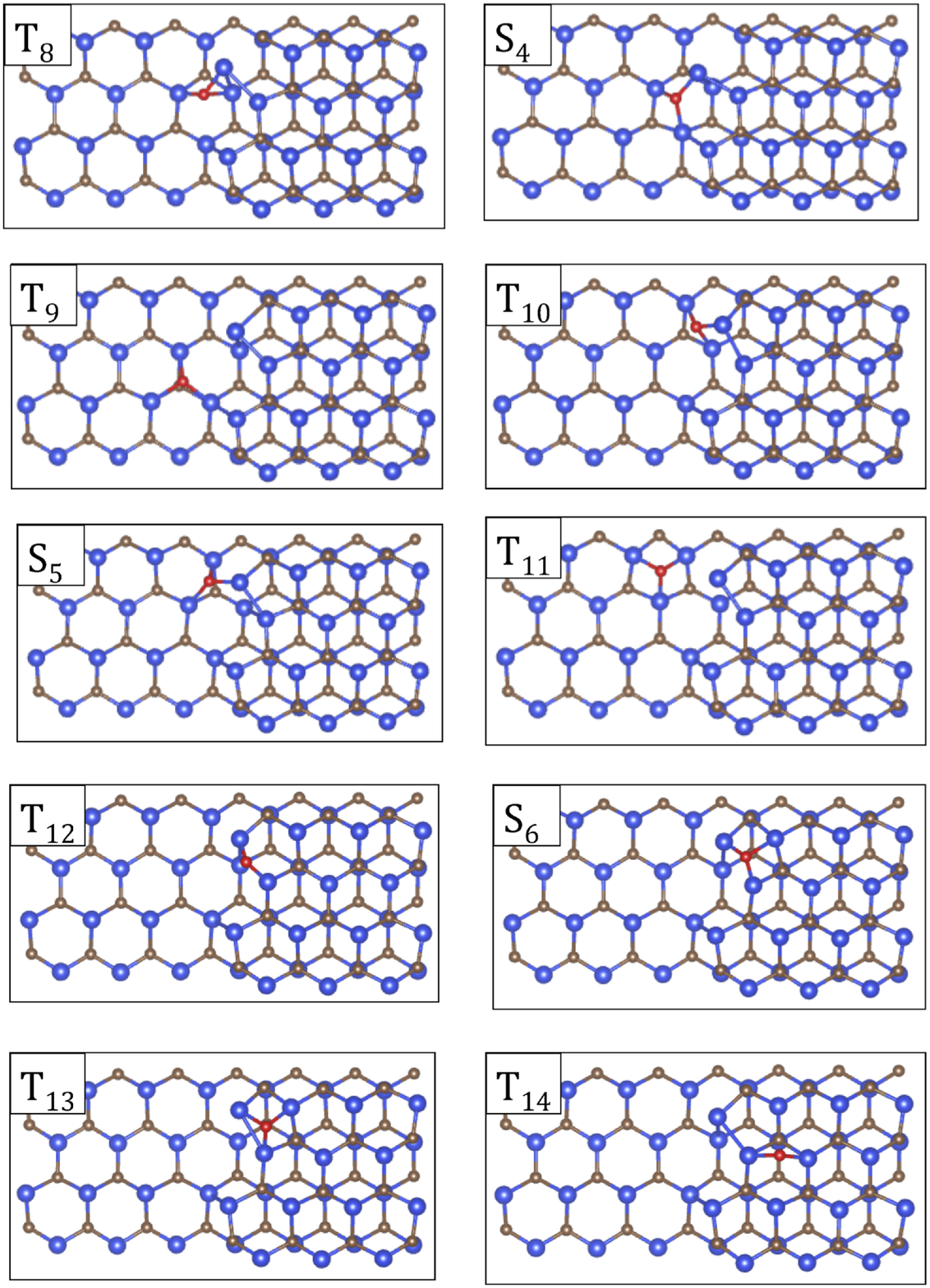}
\caption{Metastable ($S_i$) and transition-state ($T_i$) geometries along the total energy profiles of Fig.~\ref{1stC-dE} in which 
an edge C is desorbed onto T4 or H3 sites.
The moving C is evidenced by a small red sphere in each panel.}\label{1stC-S-T}
\end{figure}

\begin{table}
 \caption{Calculated energy barriers [eV] for the desorption of an edge Si or C to the 
T4 or H3 site on the lower (@L) or the upper (@U) terrace.
}\label{1st-dE}
  \begin{center}
  \begin{tabular}{l|ccc} \toprule
   \multicolumn{4}{c}{Energy barrier [eV]} \\ \hline \hline
                            & init.${\rightarrow}S_{1}$ & $S_{1}{\rightarrow}S_{2}$(=H3@L) & $S_{2}{\rightarrow}$T4@L \\
  Si: H3@L$\rightarrow$T4@L &  1.00                     &  1.03                     &  0.60 \\ \hline
                            & init.${\rightarrow}S_{3}$ &  & $S_{3}{\rightarrow}$T4@L  \\
  Si: T4@L                  &   2.27                    &  & 0.32                      \\ \hline
                            & init.${\rightarrow}$T4@U  &  & \\
  Si: T4@U                  &   3.69                    &  & \\ \hline
                            & init.${\rightarrow}$H3@U  &  & \\
  Si: H3@U                  &   2.70                    &  & \\ \hline \hline
                            & init.${\rightarrow}S_{4}$ &  & $S_{4}{\rightarrow}$T4@L \\
  C: T4@L                   &   2.86                    &  & 0.44 \\ \hline
                            & init.${\rightarrow}S_{5}$ &  & $S_{5}{\rightarrow}$H3@L \\
  C: H3@L                   &   1.78                    &  & 0.30 \\ \hline
                            & init.${\rightarrow}S_{6}$ & $S_{6}{\rightarrow}S_{7}$(=T4@U) & $S_{7}\rightarrow$H3@U \\
  C: T4@U$\rightarrow$H3@U    &   2.19                    &  0.50                     &  1.14 \\ \hline
 \end{tabular}
 \end{center}
\end{table}

We have identified four distinct pathways for the edge Si atom to be desorbed and migrate toward the terrace sites. The obtained energy barriers overcoming the transition states are shown in Table~\ref{1st-dE}. 
The calculated barriers turn out to be 1.03, 2.27, 3.69, and 2.70 eV for T4@L via H3@L, T4@L, T4@U, and H3@U reactions, respectively.
From this, the most probable process for the desorption of an edge Si atom is one toward T4@L via H3@L with the rate-determining energy barrier of 1.03 eV.
We have also identified the reaction pathways for the edge Si desorption at the SB step with the hexagonal upper and cubic lower terraces. We have obtained the energy barriers of 1.27 eV, 3.84 eV and 2.62 eV for the desorption toward T4@L via H3@L, T4@L and T4@U, respectively, indicating that the desorption toward the lower terrace is again energetically favorable.

Performing the same analysis done for the Si desorption, the obtained total energy profiles for the reaction pathways in which an edge C atom is desorbed and migrates onto 
the T4 or H3 site 
on the terraces are summarized in Fig.~\ref{1stC-dE}. The abscissa is the distance in 3$N_{\rm atom}$-dimensional space along the determined reaction pathway, as is in Fig.~\ref{1stSi-dE}. We have identified three reaction pathways: i.e., the paths toward the T4 site at the lower terrace (T4@L), the H3 site at the lower terrace (H3@L) and the H3 site at the upper terrace (H3@U) via the T4 site at the upper terrace (T4@U). All metastable and transition-state geometries labeled as $T_{i}(i=8,\cdots,14)$ and $S_{i}(i=4,\cdots,7)$ in Fig.~\ref{1stC-dE} are shown in Fig.~\ref{1stC-S-T}.

In the path toward T4@L,
$T_{8}$, $S_{4}$, and $T_{9}$ are the relevant transition and metastable states. From the initial position at the step edge, a single C atom is desorbed toward a metastable structure $S_{4}$, which is basically an H3 site on the lower terrace. In the reaction H3@L in which a C atom is desorbed to the H3 site at the lower terrace, we have found that the reaction takes place through the transition state $T_{10}$, the metastable $S_5$, the transition state $T_{11}$ and finally H3@L.

For the reaction of the edge C atom being desorbed and migrating to the upper terrace, we have found a reaction pathway where the C atom migrates to the T4 site and then continues to move the H3 site of the upper terrace: T4@U$\rightarrow$H3@U.
This reaction takes place passing through the $T_{12}$, $S_6$, $T_{13}$, $S_7$, $T_{14}$ and finally H3@U geometries. The $S_7$ here is identical to T4@U. 
The transition states $T_{12}$ and $T_{14}$ present twofold coordinated C sites. In view of this reduced coordination number,
 these geometries are indeed transition states.

\begin{figure*}
\includegraphics[width=.9\linewidth]{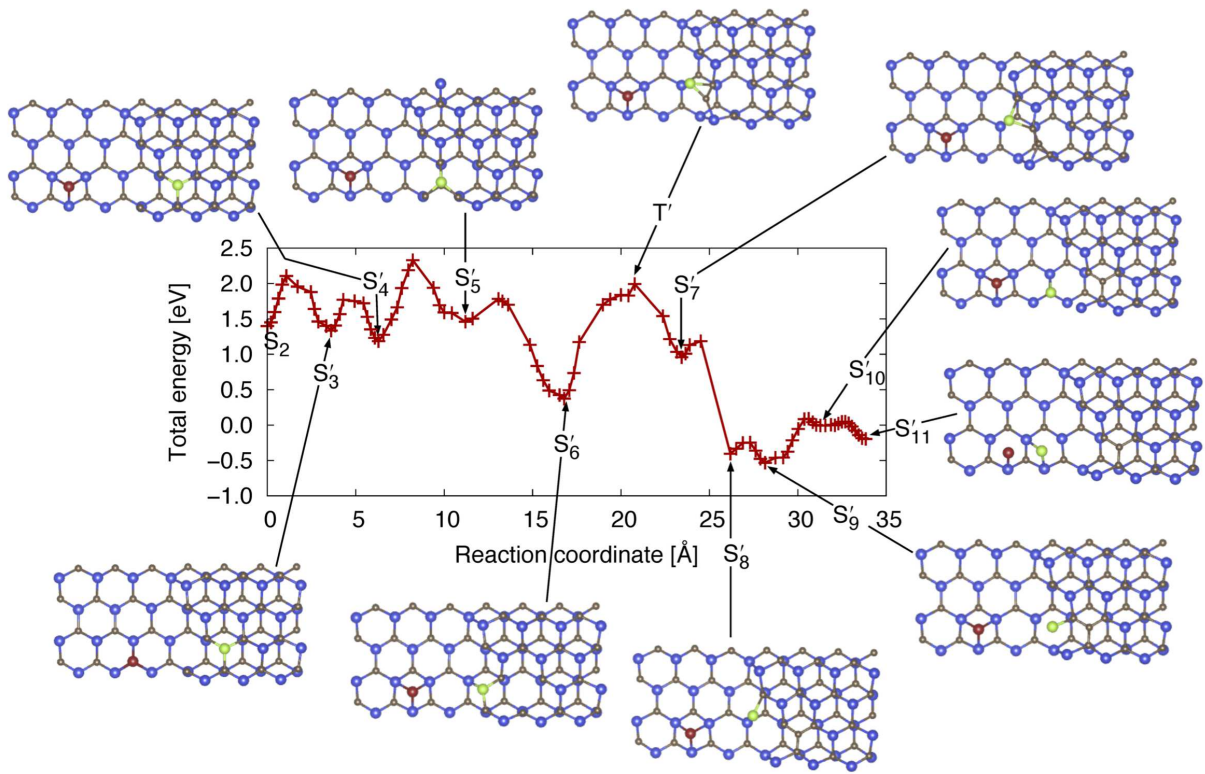}
\caption{Reaction barriers and relevant geometries of all metastable states and the main transition 
states along the reaction for a subsequent Si desorption.
The first Si atom desorbed and the second departing Si atom 
are shown as large red and green spheres, respectively.}\label{c-Si2}
\end{figure*}

\begin{figure}
\includegraphics[width=.98\linewidth]{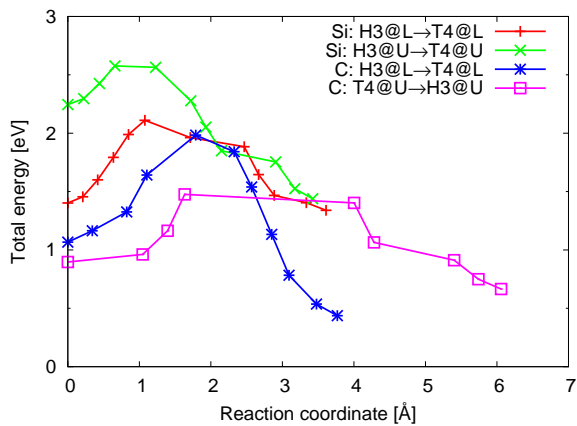}
\caption{Calculated reaction barriers for Si and C migration on the lower and upper terraces.}\label{diffusion}
\end{figure}

We have identified three distinct pathways for the edge C atom to be desorbed and migrate toward the terrace sites. The obtained energy barriers overcoming the transition states are shown in Table~\ref{1st-dE}. The calculated barriers turn out to be 2.86, 1.78 and 2.19 eV for 
T4@L, H3@L, and T4@U$\rightarrow$H3@U reactions, respectively. The barriers are substantially larger than those for the Si desorption. We have also calculated the reaction paths for a SB step with hexagonal upper and cubic lower terraces. The calculated barriers for the C desorption toward T4@L, H3@L and T4@U$\rightarrow$H3@U are 1.78 eV, 2.08 eV and 1.96 eV, respectively. The value are comparable with those for the cubic upper and hexagonal lower terraces.

Our calculations for the reaction barriers for Si and C atoms from the step edges have unequivocally reveals that the desorption of the edge Si to the H3 site on the lower terrace, characterized by the barrier of 1.03 eV, is the most probable atomic process in the edge-atom desorption. However, considering the endothermicity we have found of the whole reaction, this single Si desorption may not be a whole story but an important trigger for subsequent exothermic reactions.

\subsection{Subsequent desorption from the step edge}

In the preceding 
subsection, we have identified the reaction pathways for single-atom desorption from the SB step to suitable terrace sites and found that the energy barrier for Si desorption is lower than that for C desorption. This is indicative of the selective desorption of Si from SiC. However, our calculations show that the reaction is endothermic.
It seems then exothermic reactions, if existing, for the Si desorption are still escaping our investigation.

In the search for a possible exothermic reaction pathway in which the final state is energetically located at a lower value with respect to the initial SB step structure, we have considered a subsequent Si desorption from the structure illustrated in Fig.~\ref{fin-Si} (b), namely $S_{2}$ labeled in Fig.~\ref{1stSi-dE}, since it is the lowest in energy among the desorbed structures and accessible with the lowest energy barrier of 1.03 eV. By continuing our simulations from this configuration, we have found the geometry $S'_{11}$ as 
in Fig.~\ref{c-Si2}. This structure is indeed lower in energy than the initial SB-step structure by as much as 0.20 eV. The total energy profile and associated energy barriers between $S_{2}$ and $S'_{11}$, as provided by our computational approach, are shown in Fig.~\ref{c-Si2}.

Starting from $S_{2}$ in which the first Si atom is located at an H3 site on the lower terrace, this Si migrates to an adjacent H3 site along the reaction from $S_{2}$ to $S'_{4}$ via a T4 site labelled as $S'_{3}$ in Fig.~\ref{c-Si2}. From $S'_{4}$, a second Si atom migrates from the upper terrace to a metastable site indicated as $S'_{5}$ on the step edge, then it protrudes from the edge onto an adjacent lower H3 site $S'_{6}$. $S'_{6}$ is a local minimum carrying one Si monovacancy on the upper terrace with one Si adatom on the lower terrace. From $S^{\prime}_6$, other metastable $S^{\prime}_7$ structure appears after passing through the transition state $T^{\prime}$.
The largest barrier during the whole process is 1.62 eV and appears at $T^{\prime}$ between the intermediate steps $S'_{6}$ and $S'_{7}$. 
The structure $T^{\prime}$ is characterized by the formation of a new C--C bond. It is noteworthy that there is substantial energy decrease upon the structural change from $S^{\prime}_7$ to $S^{\prime}_8$. This is due to the increase of the number of C--C bonds in the system. More precisely, this number is two in $S'_{7}$ and three in $S^{\prime}_8$ or $S'_{9}$. The position of the second Si is different in $S'_{8}$ and $S'_{9}$, leading to a slight energy difference between them. From $S'_{9}$ to $S'_{11}$, the second Si migrates from the step edge to an H3 site on the lower terrace. 

The energy barrier (1.62 eV) is higher than the barrier of 1.03 eV by which the first Si atom undergoes desorption and migrates to the lower terrace via an H3 site. Nonetheless, the barrier of 1.62 eV is lower than other desorption rate-determining barriers (see Table~\ref{1st-dE}) calculated in section \ref{ssec:1Si}.

To complete our inspection, we have also considered the migration of Si and C atoms on the lower and upper terraces. We have considered the migration starting after the desorption process with relatively lower energy barriers, namely H3@L and H3@U for Si, and H3@L and T4@U for C. Fig.~\ref{diffusion} shows the related total energy profiles for reactions in which one Si or C atom migrates from an H3 or a T4 site. The Si migration barriers estimated from these profiles are 0.71 and 0.33 eV on the lower and upper terrace, respectively. The C migration barriers are always higher than those of Si both on the lower and upper terrace: 0.92 and 0.58 eV on the lower and upper terrace, respectively. In each transition state which connects the initial and final metastable states both presenting a threefold coordination,  the migrating atom is found to be twofold coordinated. The highest Si migration barrier (0.71 eV) is lower than the first Si desorption to the lower terrace (1.03 eV).

From these results, we conclude that a single Si desorption from the step edges triggers the exothermic reaction in which two Si atoms are selectively desorbed from the step edge and that the desorbed Si atoms migrate subsequently on the lower terrace. The exothermicity of this two-Si desorption from the step edge comes from the formation of C -- C bonds near the edges. This is certainly a seed of graphene flakes.

\section{conclusions}\label{sum}

We have performed density-functional total-energy electronic-structure calculations that clarify microscopic mechanisms 
of graphene formation on SiC(0001) surfaces at its initial stage. 
We have explored favorable reactions for the desorption of either Si or C atom 
from the step edges, which are commonly observed on the surfaces of the most stable polytype 4H-SiC, by determining the desorption and the subsequent migration pathways and calculating the corresponding energy barriers for the first time. We have found that the energy barrier for the desorption of an Si atom at the step edge and the subsequent migration toward stable terrace sites are lower than that of a C atom by 0.75 eV. This is a clear evidence of the selective desorption of Si from the SiC surface which is essential for the graphene formation on the surface. However, we have also found that this single-atom desorption is an endothermic reaction, inferring that unidentified exothermic reactions are hidden. We have indeed found that the subsequent second Si desorption is such exothermic reaction. Our density-functional calculations have unequivocally revealed that two Si atoms are desorbed from the step edge and migrate toward stable sites on the terrace with the energy gain of 0.2 eV compared with the stable clean stepped surfaces. This exothermicity comes from the energy gain caused by the bond formation between C atoms being left at the step edge. This structural outcome is the seed of the graphene flakes. We have also found that the energy barriers for the single-Si and two-Si desorption reactions are about 1.0 -- 1.5 eV, reflecting the structural characteristics of the transition state, i.e., the two-fold coordinated desorbed atom. 
It is thus highly likely that the subsequent Si desorption also costs this amount of energy per atom. On the other hand, the growth of carbon bond network near the step edge provides larger energy gain with the increase in the number of departing Si atoms or equivalently the number of the lonely C atoms. This energetics is the microscopic reason for the graphene formation on the SiC surface.

\begin{acknowledgments}
We are grateful to Li Han and Yu-ichiro Matsushita for insightful discussions.
The work was partly supported 
by the project for Priority Issue ``Creation of new functional devices and high-performance materials to support next-generation industries" to be tackled by using Post-K Computer, conducted by Ministry of Education, Culture, Sports, Science and Technology, Japan. 
Computations were performed at ISSP, The University of Tokyo, at the COMA system of University of Tsukuba and at RCCS, National Institutes of National Sciences.
M.~B.~acknowledges LaBex "Nanoparticles Interacting with their Environment" ANR-11-LABX-0058\_NIE.
\end{acknowledgments}

\providecommand{\noopsort}[1]{}\providecommand{\singleletter}[1]{#1}%

\end{document}